\documentstyle[aps,prl,epsf,epsfig,floats,twocolumn]{revtex}
\begin{document}
\draft
\twocolumn[\hsize\textwidth\columnwidth\hsize\csname @twocolumnfalse\endcsname

\title{Landau Theory of the Finite Temperature Mott Transition}
\author{G. Kotliar$^{\ast }$, E. Lange$^{\ast }$,   and M.J. Rozenberg$^{+}$}

\address{ $^{\ast }$Serin Physics Laboratory, Rutgers University, 136
Frelinghuysen Road, Piscataway, New Jersey 08854, USA}

\address{ $^{+}$Departamento de F\'{\i}sica, FCEN, Universidad de Buenos Aires,
Ciudad Universitaria Pab.I, (1428) Buenos Aires, Argentina.}

\date{\today}
\maketitle
\begin{abstract}

In the context of the dynamical mean-field theory of the Hubbard
model, we identify microscopically an  order parameter for the
finite temperature Mott endpoint. We derive a Landau functional
of the order parameter. We then use the order parameter theory to
elucidate the singular behavior of various physical quantities 
which are experimentally accessible.

\end{abstract}

]

When the strength of the electron-electron interaction $U$ is
increased compared to the bare bandwidth $2D$, a metal-insulator
transition (MIT) occurs \cite{Mott}. This phenomenon, known as the
Mott transition, can take place in the absence of magnetic long-range
order, and is still an outstanding problem in condensed-matter
physics. From a theoretical point of view, a difficulty is
the absence of an  obvious order parameter to systematize the 
critical behavior of the observable quantities  when the metal 
insulator transition is not accompanied by the onset of magnetic 
long range order. These issues are experimentally
relevant to  
 systems such as V$_2$O$_3$ and Ni(Se,S)$_2$
and  are the subject of intensive experimental study. \cite{Imada:1998}
 
In recent years, great progress has been made by using the dynamical
mean-field theory (DMFT)
\cite{Georges:1996}. This framework 
describes  both paramagnetic metallic and paramagnetic insulating 
phases. The $U$-$T$ phase diagram ($T$ is the temperature) of the 
frustrated Hubbard model in
the limit of large lattice coordination is qualitatively similar to
that of the V$_2$O$_3$ and Ni(Se,S)$_2$ systems: A first-order
phase-transition line ends in a second-order critical point,
henceforth referred to as the Mott critical point, which is the main 
focus of this letter. We will use this framework to address the 
fundamental questions raised in the previous paragraph.
 
There are two earlier qualitative ideas as to what  should be the
order parameter to describe the physics around the
finite temperature  Mott  point.   One idea is to connect the
order parameter to the notions of  "metallicity" or coherence.
It can be traced back to the early paper of Brinkman and Rice
\cite{Brinkman:1970} and is captured in a slave boson formalism
where the metallic state has a non zero expectation value of a 
Bose field which describes the coherent propagation of one particle 
excitations. \cite{Kotliar:1986} 
In a very different picture, Castellani {\it et al.} 
viewed the metal as a liquid rich in doubly occupied sites, and the 
insulator as a liquid with few doubly occupied sites. The metal to 
insulator transition  is viewed as a condensation of doubly occupied 
sites, and the order parameter is related to the 
Blume-Emery-Griffith model \cite{Castellani:1979}. The Landau 
approach presented here provides a synthesis of these ideas.
It  bridges  naturally between a picture based on
one particle excitations and a picture based on
local collective  excitations (or double occupancies).
In agreement  with  Castellani {\it et al.} we find that the
Mott transition has indeed an Ising-like character.
On the other hand, we obtain a complementary
description in terms of the one particle spectral function
reminiscent of the slave boson picture.
A simple and clear description of the
critical behavior near the critical point emerges. It allows
us to systematically derive the critical behavior of
any observable quantity and to relate its non analytic dependence on 
$T$ and $U$ to that of the order parameter. Our results should be 
also of help in resolving some controversies on the solution of
the Hubbard model in infinite dimensions
\cite{Schlipf:1999,Rozenberg:1999} by providing a theoretical
framework in which to analyze numerical results on the finite
temperature Mott transition. It can also be used to analyze
results of photoemission and optical conductivity   experiments.

For simplicity, we focus on the single-band Hubbard model at
half-filling,
\begin{equation}
 \hat{H}=-\frac{t}{\sqrt{z}}\sum_{\langle ij\rangle\sigma}
	c_{i\sigma}^{+}c_{j\sigma} 
 +U\sum_i \hat{n}_{i\uparrow}\hat{n}_{i\downarrow}.
\label{hubbard}
\end{equation}
The first term describes the hopping between nearest neighbors on a
lattice with coordination number $z$. The corresponding half bandwidth
is our unit of energy, $D=2t=1$. The second term is an on-site
interaction suppressing double occupancies by imposing an energy cost
$U$ on each one. In the limit of infinite dimensions,
$z\rightarrow\infty$, this model can be mapped onto a single-impurity
Anderson model (SIAM) supplemented by a self-consistency condition.
We adopt a semicircular density of states, which is realized on
the Bethe lattice.
The dynamical 
mean-field equations can be obtained by differentiating the Landau 
functional
\begin{equation}
F_{\mbox{\scriptsize
 	LG}}[\Delta]=-T\sum_n\frac{\Delta(i\omega_n)^2}{t^2}
	+F_{\mbox{\scriptsize imp}}[\Delta],
\label{f_landau}
\end{equation}
with respect to the hybridization function $\Delta(i\omega_n)$ of the
SIAM, which has the meaning of a Weiss field. $i\omega_n$ are
fermionic Matsubara frequencies, while $F_{\mbox{\scriptsize
imp}}[\Delta]$ is the free energy of the SIAM, given by the action
$S_{\mbox{\scriptsize imp}}=S_{\mbox{\scriptsize loc}}[\Delta=0]
+\sum_{\sigma,n}f_{\sigma}^+(i\omega_n)
\Delta(i\omega_n)f_{\sigma}(i\omega_n)$.
Here, $S_{\mbox{\scriptsize loc}}[\Delta=0]$ is the action of the
local $f$ level with the hybridization set to zero. The first term in
Eq.\ (\ref{f_landau}) is the cost of forming the Weiss field
$\Delta(i\omega_n)$ around a given site, while the second one is the
free energy of an electron at this site in the presence of the Weiss
field. Using the Green's function of the SIAM, 
$G(i\omega_n)=(1/2T)\delta F_{\mbox{\scriptsize imp}}/\delta
\Delta(i\omega_n)$, the mean-field equation reads
\begin{equation}
 \frac{t^2}{2T}\frac{\delta F_{\mbox{\scriptsize LG}}[\Delta]}
	{\delta\Delta(i\omega_n)}=t^2G(i\omega_n)[\Delta,\alpha]
	-\Delta(i\omega_n)=0.
\label{mean-field} 
\end{equation}
Here, $\alpha=(U,T)$ comprises the control parameters. This Landau approach 
was used to describe the energetics  of the Mott
transition at zero temperature \cite{gabi}. 
We will show that near the 
finite temperature Mott point,    the Weiss field 
has a singular dependence which can be parametrized
by a single number which assumes the role of an 
effective order parameter for this transition.

As in Landau theory, we {\it assume} that a finite temperature
transition exists, and  {\it derive} a complete description
of  the critical 
behavior  near  the transition  as follows:
First, we expand the mean-field equation (\ref{mean-field}) around the
critical point, $\alpha_c=(U_c,T_c)$, up to third order in the
deviation of the hybridization function from its value at the critical
point, $\delta\Delta=\Delta(\alpha_c+\delta\alpha)-\Delta(\alpha_c)$,
and to first order in $\delta\alpha=(U-U_c,T-T_c)$. This expansion is 
well-behaved because the impurity model {\it at finite temperatures}
depends smoothly on $\alpha$ 
and $\delta \Delta(i\omega_n)$.
In order to carry out this expansion it is convenient to define a 
fluctuation matrix
\begin{equation}
 M_{nm}=\frac{t^2}{2T}\left.\frac{\delta^2F_{\mbox{\scriptsize
   LG}}[\Delta]}{\delta\Delta(i\omega_n)\delta\Delta(i\omega_m)}
   \right|_{\mbox{\scriptsize critical point}}
\label{fluct_mat}
\end{equation}

$M_{nm}$ has the form $-{\delta_{nm}}+K_{nm}$, where
$K_{nm}$ is  the Fourier transform of a
kernel  $ K(\tau, \tau')$
which is proportional to  the connected correlation function of an operator
$O(\tau)={{\int_0}^{\beta} du {f^+}(u+\tau)  {f}(u)}$,
$<O(\tau) O(\tau')> -<O(\tau)>< O(\tau')>$
where the average $<>$ 
is  calculated with the action of an  Anderson  impurity model. 
It is well known that the  correlation functions of the Anderson
impurity model are  {\it bounded},
and therefore the Kernel $K$ is square 
integrable $ {\int_0}^{\beta} {\int_0}^{\beta} d\tau  d\tau'
|K(\tau, \tau')|^2 < \infty $. Therefore it $K_{nm}$ is a Fredholm operator
which  and has a {\it discrete} spectrum of eigenvalues which
we labeled by the index  $l$.  \cite{footnote}

At half-filling, particle-hole symmetry guarantees that the order
parameter $\Delta(i\omega)$ is odd and wholly imaginary. Accordingly,
the fluctuation matrix
is real and symmetric and  has real eigenvalues $m_l$
belonging to eigenvectors $\phi_l(i\omega_n)$ which can be chosen to
be purely imaginary and to form an orthonormal basis. The critical
point, in this description of the problem, is signaled by the
appearance of a single zero eigenvalue, $m_0=0$, which indicates the
occurrence of a simple bifurcation.

Next, we represent $\delta\Delta$ in the 
eigenbasis of the matrix (\ref{fluct_mat}),
$\delta\Delta(i\omega_n)=\sum_l\eta_l\phi_l(i\omega_n)$, where all
$\eta_l$ are real. By projecting the mean-field equation 
(\ref{mean-field}) onto the eigenbasis $\phi_l$, we obtain an equation 
of the form
\begin{eqnarray}
 &m_l\eta_l+F^{(0)}_l[\{\eta_{j\ne0}\}]
	+F^{(1)}_l[\{\eta_{j\ne0}\}]\eta_0&
\nonumber\\
	&+F^{(2)}_l[\{\eta_{j\ne0}\}]\eta_0^2
	+F^{(3)}_l\eta_0^3=0&,
\label{expansion}
\end{eqnarray}
which holds for all $l$. $F^{(0)}_l$ is of order $\delta\alpha$. 
$F^{(1)}_l$ and $F^{(2)}_l$ have Taylor expansions in the 
$\eta_{j\ne0}$, where $F^{(1)}_l$ starts with the linear order. We 
solve Eq.\ (\ref{expansion}) iteratively for all $\eta_{l\ne0}$ to 
obtain $\eta_{l\ne0}=a_l+c_l\eta_0^2+d_l\eta_0^3$. Here, $a_l$ is of 
first order in $\delta\alpha$, (which  assures us that  the leading
singular dependence of the spectral function is proportional to $\phi_0$)   
further corrections have the form $b_l\eta_0$ 
with $b_l$ also of order $\delta\alpha$. By inserting this expression
into the $l=0$ case of Eq.\ (\ref{expansion}), we derive an effective 
equation for the zero-mode amplitude $\eta_0$. We can think of 
$\eta_0$ as the soft mode near the transition and the $\eta_{l\neq 0}$ 
as massive modes. The elimination of the massive modes renormalizes 
the coefficients of the effective action for the soft mode. In the 
resulting cubic equation for $\eta_0$, we eliminate the quadratic term 
by shifting $\eta_0$ by an appropriately chosen linear function in 
$\delta\alpha$, 
$\eta=\eta_0+\mbox{const}_1\times(T-T_c)+\mbox{const}_2\times(U-U_c)$.
Close to the critical point, $\eta$ and $\eta_0$ are dominated by non 
analytic terms and are therefore essentially equal. We thus obtain an 
equation of state without quadratic term in $\eta$:
\begin{equation}
 p\eta+c \eta^3=h.
\label{eqofstate}
\end{equation}
Here, all quantities are real.

As in Landau
theory,  a microscopic calculation of
the  Landau coefficients  (p,c,h) is difficult.
However we can extract exact information about
the critical behavior from the knowledge
that they are smooth
functions of the control parameters, i.e.  $c$ is finite at the
critical point, whereas $p$ and $h$ are linear functions of
$\delta\alpha$, $h=h_1(U-U_c)+h_2(T-T_c)$ and
$p=p_1(U-U_c)+p_2(T-T_c)$. As a consequence, $\eta$ has a singular 
dependence on $U$ and $T$ near the critical point. At $U=U_c$, 
and for $T$ near $T_c$,
\begin{equation}
\eta(U_c,T)\simeq \mbox{sign}(h_2/c)
\mbox{sign}(T-T_c)|T-T_c|^{1/3}.
\end{equation}

The mean-field equation (\ref{eqofstate}) describes the Mott transition 
close to the critical point in terms of the order parameter $\eta$. In 
this form, the analogy with the liquid gas transition is evident. The 
Mott transition takes place on the line in the $U$-$T$ plane where $h$ 
vanishes and the system has full Ising symmetry. The critical point, 
$(U_c, T_c)$, divides this line into two half-lines. On the half-line 
where $T<T_c$, there are two solutions, $\eta=\pm\sqrt{|p/c|}$. 
We will see later that $\eta$ parametrizes the strength of the 
quasiparticle resonance of the single-particle spectrum (see Fig.\ 
\ref{fig2}). A positive or negative "field" $h$ increases or decreases 
this component of the spectral function, respectively. The field $h$ 
decreases when $U$ or $T$ is increased, because either increase 
eliminates the metallic coherence and thus reduces the value of $\eta$. 
We have used the sign convention 
whereby is positive.

We now turn to  various consequences of our construction.
{}From Eq.\ (\ref{eqofstate}), we can obtain the {\it shape}
of the coexistence region near the critical point, where
two solutions of the mean field equations coexist.
It is centered
symmetrically about the $h=0$ line, and its width along
$T=\mbox{const}$ lines, $\Delta U$, scales with $(T_c-T)^{3/2}$. The
constant of proportionality is given by
$(4/\sqrt{c}\,|h_1|)[(p_2-p_1h_2/h_1)/3]^{3/2}$. 

An important quantity which is measured in numerical simulations
is the double occupancy. It is connected to our order parameter $\eta$
as follows: $\langle d\rangle=(T/U)\sum_n\{[(i\omega_n+\mu)
G(i\omega_n)-1]e^{i\omega_n0^+}-t^2G(i\omega_n)^2\}=\langle
d\rangle_c+c_1^{(d)}\eta+c_2^{(d)}\eta^2$. In this expansion about the 
critical point, we have only retained the leading
and next to leading  nonanalytic terms 
responsible for the critical behavior. The susceptibility 
$\chi=\partial\langle d\rangle/\partial U$ diverges 
at the critical point.  For example:
\begin{equation} 
\chi(U,T_c)\simeq 
(c_1^{(d)}/3)\mbox{sign}(h_1/c)|h_1/c|^{1/3}|U-U_c|^{-2/3}.
\end{equation}
The double occupancy is related to the magnetization by the identity 
$\langle(n_{\uparrow}-n_{\downarrow})^2\rangle=1-2\langle d\rangle$.
The magnetic response will therefore also exhibit nonanalytic 
dependences on the control parameters.

\begin{figure}
  \epsfxsize=3.5in
  \epsffile{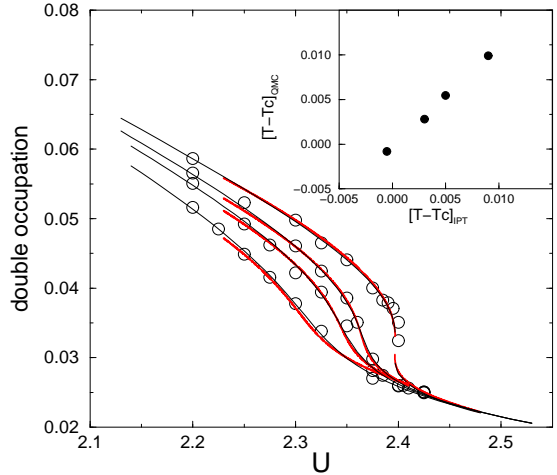}
  \caption{
Double occupation  $\langle d\rangle$ as a
function of $U$ for different temperatures. The thin black lines denote
IPT results for $T_{\mbox{\scriptsize IPT}}=0.0469, 0.05, 0.052, 0.056$
(top to bottom). The thick red lines
are a fit to the IPT data using the LG theory. The circles are
QMC data obtained at $T_{\mbox{\scriptsize QMC}}= 1/40, 1/35, 1/32, 1/25$
{\protect \cite{Rozenberg:1999} }. 
The IPT results where shifted by a constant -0.07
along the $U$ axis and by -0.003 along the $\langle d \rangle$ axis.
The curves for the 3 larger temperatures are above $T_c$ and the 
lowest temperature ones (2 branches) are just below.
The inset shows the scaling of the {\it reduced} temperatures
$[T-T_c]_{\mbox{\scriptsize QMC}}$ versus
$[T-T_c]_{\mbox{\scriptsize IPT}}$.
}
\label{fig1}
\end{figure}

\begin{figure}
	\epsfxsize=3.5in
 	\epsffile{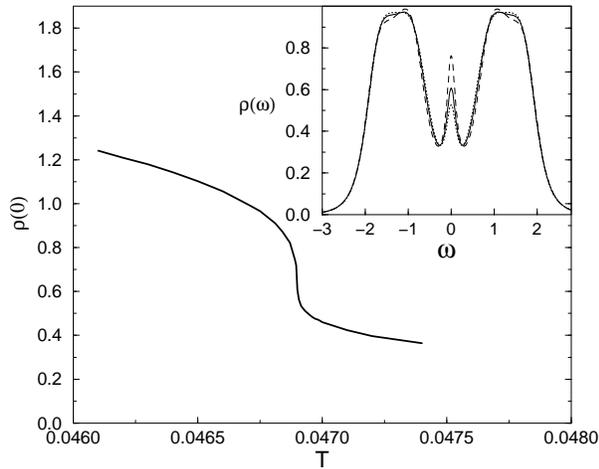}
	\epsfxsize=3.5in
	\caption{The density of states at the Fermi energy $\rho(0)
\equiv A_0$
as a function of temperature
in the critical region ($U=2.46316 \approx U_c$). 
The singular behavior of the slope
at $T_c \approx 0.046897$ can be clearly appreciated. The inset shows
the variation of the 
spectral function for $U\simeq U_c$ in the vecinity
of $T_c$: dashed line for $T-T_c/T_c=-0.00025$, 
solid line for $T-T_c/T_c=0.00006$ and
dotted line for $T-T_c/T_c=0.00049$. {\protect \cite{caveat}}
}
\label{fig2}
\end{figure}

There has been several numerical studies of the
finite temperature 
Mott transition in this model. The Landau approach predicts
the functional dependence of various quantities near the transition,
and therefore the expressions derived in this paper,  are useful
for interpreting the numerical work.
To illustrate how
our approach  sheds  new light on previously obtained numerical data
we compare 
in Fig.\ \ref{fig1}  the results for the double occupancy 
$\langle d \rangle$ obtained within the IPT and QMC calculations with 
$\Delta\tau=0.5/D$, after  carrying out the shifts and the rescaling
described in the figure caption.
Within the statistical errors of the QMC calculation, the agreement is
excellent.  This  surprising result is consistent with the  Landau 
theory:
different  approximations for the solution of the impurity 
model reduce to the same Landau theory near the critical point, but with 
different values of the Landau coefficients. Therefore, with a suitable 
rescaling, the results near the critical point should agree with each
other, and with a fit based on the Landau theory which is shown
in the red line in figure 1. 

Small changes in the values of $\Delta\tau$  result in shifts of
$U_c$, $T_c$, and $\langle d \rangle$ at criticality, but 
does not change the form of
the critical behavior.
We  also note  that the critical slowing down which
has been observed in the iterative solutions of the mean field
equations are a direct consequence of the presence of the soft mode $\eta$
described in the Landau approach.

{}From  our construction it is clear that 
$\eta$ provides the leading non analytic behavior of the Weiss field.
In order to get a better feeling for its physical significance we
have to understand how it can be probed experimentally. 
Since the order parameter is closely related to the amplitude
of the quasiparticle peak, photoemission is an ideal tool to probe the
temperature and pressure dependence of the order parameter near the
critical point. This experimental technique, in the angle integrated 
mode, would also measure the convolution of the Fermi function with 
the analytically continued eigenfunction of the zero mode, 
$\mbox{Im}\phi_0(i\omega_n = \omega - i \delta)$. 
To visualize the shape of the spectral function near the critical point
we must resort to calculations based on analytic methods such as IPT.
The inset of Fig.\
\ref{fig2} shows the spectral function very near the critical point, computed
within the IPT.
It illustrates how the compromise between metallic and insulating
features is realized. A finite $\eta$, depending on its sign, adds or
subtracts spectral weight to the coherent low energy feature immersed
in a constant backround in between the Hubbard bands.
The zero mode is seen to 
affect mainly the low-energy part of the spectrum, 
which determines whether the system is metallic or insulating. The  
strong temperature dependence has been noticed in  previous theoretical 
and experimental studies. \cite{Matsuura:1996} Its origin and 
connection to an order-parameter description of the Mott transition, 
however, had not been recognized until now.
In the main panel of Fig.\ \ref{fig2} we display the height of the 
quasiparticle peak 
$A_0=i\Delta(i0^+)/\pi t^2$,
for $U\simeq U_c$, 
as a function of temperature in the vicinity of $T_c$.
The rapid variation seen in the figure is consistent with  the form
$A_0=A_{0c}+c_1^{(A)}\eta+c_2^{(A)}\eta^2$ 
with coefficients $ c^{(A)}_i$  independent of  U and temperature.

Optical techniques are probably  the  best tool  available to test 
the predictions of our theory. 
For instance, one may consider the integral of the optical
conductivity up to some cuttoff,
$N_{\mbox{\scriptsize eff}}(T)$. Since the optical conductivity in 
infinite dimensions is directly expressed in terms of the 
single-particle Green's function, $N_{\mbox{\scriptsize eff}}(T)$ must 
also exhibit the singular temperature dependence near the transition. 
We would therefore 
expect  the  temperature variation  of this  quantity to be most 
visible for a relatively small cuttoff, 
displaying a rapid variation with $T$ similarly as for $A_0$.
Since the singular
dependence arises from the order parameter $\eta$, it should be
possible to fit the Drude weight
by $N_{\mbox{\scriptsize eff}}(T)=N_{\mbox{\scriptsize
eff}}(T_c)+c_1^{(N)}\eta(T)+c_2^{(N)}\eta^2(T)$.
$N_{\mbox{\scriptsize eff}}(T)$ has recently been measured 
in NiS$_{2-x}$Se$_x$ \cite{Takagi:1999},   
the observed strong temperature dependence of
the effective number of carriers is consistent with our predictions.
 
In summary, we derived an  order parameter description
of   the Mott transition
near its critical point in the $U$-$T$ plane. We showed that the
critical behavior in proximity to this point is governed by an
Ising-like Landau functional and is present in a large number
of observable quantities.
 We predict that any physical 
quantity which is sensitive to the single-particle spectrum exhibits 
singular dependences on the control parameters close to the 
finite-temperature Mott point.
The  leading non analytic
behavior of  other physical quantities can be obtained  along similar lines,
i.e. by recognizing  their coupling to the order parameter. This involves  
a few coefficients, (i.e. the $c^{(A)}$'s)
which depend on  the observable (and on
the approximation method) and, as in Landau theory, 
should be taken as parameters. The dependence on
temperature and on pressure is completely determined
from the temperature or pressure dependence of the order
parameter that follows from Eq.\ (\ref{eqofstate}).
ACKNOWLEDGMENT
This work was supported by NSF 95-29138. E.L. was  partially supported
by the Deutsche Forschungsgemeinschaft. M.J.R. acknowledges support of
Fundaci\'on Antorchas, CONICET (PID $N^o4547/96$), and ANPCYT
(PMT-PICT1855). We  thank R. Chitra for  discussions 
and D. Vollhardt for useful comments on the mansucript.


\begin{thebibliography}{10}
\bibitem{footnote}

This is only true at finite temperatures, which ensures
that the limit of integration $\beta $ is finite. At zero temperature
the spectrum of fluctuations is continuos and the transition
at a point denoted $U_{c2}$ has a very different character and was 
described 
in Ref. \cite{gabi}.
\bibitem{caveat} To accurately locate the critical point with IPT, we 
found it best to solve the DMFT equations on the imaginary 
axis. We note that closest to the $(U_c,T_c)$ point,
up to 10,000 iterations were required to find converged solutions.
\end{thebibliography}
\end{document}